\documentclass[journal=jpclcd]{achemso}

\usepackage[version=3]{mhchem} 
\usepackage[T1]{fontenc}       
\usepackage{braket}
\usepackage{amsfonts}
\usepackage{color}
\usepackage{amsmath}
\usepackage{amssymb}
\usepackage{verbatim}
\usepackage{latexsym}
\usepackage{enumerate} 
\usepackage{bm} 
\usepackage{ulem} 
\usepackage[caption=false]{subfig}
\usepackage{multirow}
\usepackage{booktabs}
\usepackage{lineno}

\author{Cory Jones}
\affiliation{Selwyn College, University of Cambridge, Grange Road, Cambridge CB3 9DQ, United Kingdom}
\author{Bo Peng}
\affiliation{Theory of Condensed Matter Group, Cavendish Laboratory, University of Cambridge, J.\,J.\,Thomson Avenue, Cambridge CB3 0HE, United Kingdom}
\email{bp432@cam.ac.uk}

\title[An \textsf{achemso} demo]
{
Boosting Photocatalytic Water Splitting of Polymeric C$_{60}$ by Reduced Dimensionality from 2D Monolayer to 1D Chain
}


\begin{document}

\begin{tocentry}

\includegraphics[width=1\linewidth]{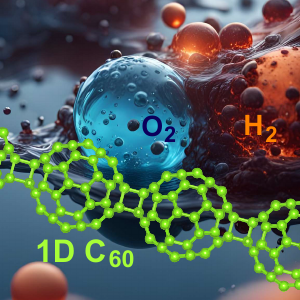}




\end{tocentry}

\begin{abstract}

Recent synthesis of monolayer fullerene networks [\textit{Nature} \textbf{2022}, 606, 507] provides new opportunities for photovoltaics and photocatalysis because of their versatile crystal structures for further tailoring of electronic, optical and chemical function. To shed light on the structural aspects of photocatalytic water splitting performance of fullerene nanomaterials, we compare the photocatalytic properties of individual polymeric fullerene chains and monolayer fullerene networks from first principles calculations. It is found that the photocatalytic efficiency can be further optimised by reducing dimensionality from 2D to 1D. The conduction band edge of the polymeric C$_{60}$ chain provides a much higher external potential for the hydrogen reduction reaction than its monolayer counterparts over a wider range of pH values, and the surface active sites in the 1D chain are two times more than those in the 2D networks from a thermodynamic perspective. These observations render the 1D fullerene polymer a more promising candidate as a photocatalyst for the hydrogen evolution reaction than monolayer fullerene networks.

\end{abstract}



In photocatalytic water splitting, water is decomposed into hydrogen and oxygen using light, which can produce hydrogen as a green energy alternative to fossil fuels\,\cite{Fujishima1972,Norskov2004,Rossmeisl2007,Zhang2007,Deak2011,Scanlon2013,Pfeifer2013,Ju2014,Mi2015,Zhang2015n,Deak2016,Chiodo2010,Li2020a}. Recently synthesised 2D fullerene networks\,\cite{Hou2022,Meirzadeh2023} hold great promise for such applications owing to the benefits of (1) a suitable band gap to generate a large amount of electron-hole pairs, (2) high carrier mobility to separate electrons and holes, 
and (3) appropriate band edges to thermodynamically drive the hydrogen evolution reaction with the help of photoexcited electrons, and the computational predictions based on such microscopic mechanisms\,\cite{Peng2022c} have been recently confirmed experimentally\,\cite{Wang2023}. Among the different structural phases of polymeric C$_{60}$ monolayers, the quasi-1D quasi-tetragonal phase (qTP1) and the tightly bound quasi-tetragonal phase (qTP2) have enhanced photocatalytic water splitting performance over the quasi-hexagonal phase (qHP). However, monolayer qTP2 C$_{60}$ is thermodynamically less stable than monolayer qTP1 C$_{60}$ at room temperature, while monolayer qTP1 C$_{60}$ tends to split into individual 1D chains because of its low dynamic and mechanical stability\,\cite{Peng2023}. Therefore, it is worth investigating the 
electronic, optical, transport and thermodynamic properties of the 1D fullerene polymer in the context of the photocatalysis of the hydrogen evolution reaction. 

The theoretical prediction and experimental synthesis of the 1D fullerene polymer can be dated back to the 1990s\,\cite{Xu1995,Springborg1995,Nunez-Regueiro1995,Marques1996}. In a 1D fullerene chain, each C$_{60}$ cage connects neighbouring cages through covalent [$2+2$] cycloadditional bonds, which can be formed between isolated C$_{60}$ molecules as a result of photo- or pressure-induced polymerisation\,\cite{Rao1993,Iwasa1994,Eklund1995}. The 1D polymeric C$_{60}$ chain can be viewed as the building blocks for a variety of ordered or disordered 1D, 2D and 3D fullerene/fullerite polymers\,\cite{Giacalone2006,Murga2015} with distinct electronic, optical and vibrational properties\,\cite{Venkateswaran1996,Rao1997,Forro2001,Makarova2001,Sun2005}. Additionally, polarons and self-trapped excitons can be formed in the 1D C$_{60}$ crystals, resulting in Jahn-Teller distortion that can be described by the Su-Shriffer-Heeger model\,\cite{Belosludov2003,Belosludov2006}. The physical and chemical properties of 1D polymeric C$_{60}$ chains can be further tuned by doping, leading to various applications such as superconductivity\,\cite{Chauvet1994,Stephens1994,Gunnarsson1997,Huq2001}. However, it is unclear whether 1D fullerene polymers can be assessed as a possible photocatalyst of the hydrogen evolution reaction. In particular, the physical and chemical implications of reducing the dimensionality from 2D to 1D C$_{60}$ on photocatalysis are still uninvestigated. 

In this work, we investigate the band structures of the 1D C$_{60}$ chain and 2D qTP2 C$_{60}$ networks using the unscreened hybrid functional, which accurately describes the effects of reduced dimensionality. Due to the similarities in their structures, a direct comparison between 1D C$_{60}$ and 2D qTP2 C$_{60}$ polymers is possible. The excitonic and transport properties of polymeric fullerene chains are also investigated to understand whether electrons and holes can be separated effectively upon photoexcitation. Finally, the free energy barrier for the intermediates in the hydrogen evolution reaction is calculated for all possible adsorption sites, which further confirms that 1D C$_{60}$ has much higher photocatalytic efficiency owing to twice the energetically favoured adsorption sites and larger surface area compared to their monolayer counterpart.



\begin{figure}
    \centering
    \includegraphics[width=0.5\linewidth]{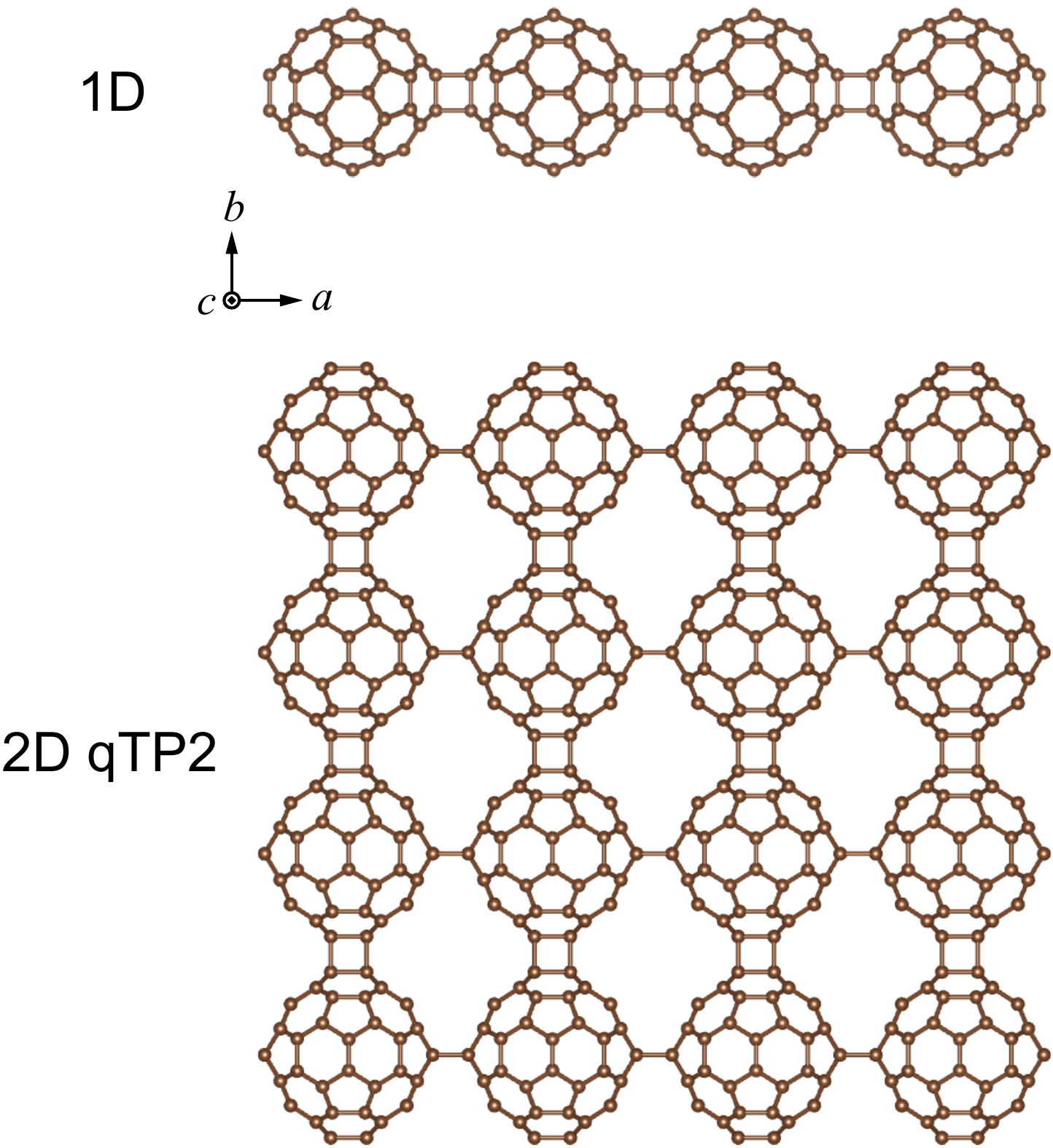}
    \caption{Crystal structures of 1D and 2D C$_{60}$ from the top view.}
    \label{crystal}
\end{figure}

Figure\,\ref{crystal} shows the crystal structures of polymeric fullerene chains and monolayer qTP2 fullerene networks. Both structures have inversion symmetry and are mirror symmetric with respect to the (100), (010) and (001) planes, with $C_2$ rotational symmetry along the [100], [010] and [001] axes. For 1D C$_{60}$, each C$_{60}$ cage is connected by the in-plane [$2+2$] cycloaddition bonds along the $a$ direction; in 2D qTP2 C$_{60}$, the C$_{60}$ cages along $a$ and $b$ are linked by the vertical and in-plane [$2+2$] cycloaddition bonds respectively. Structural relaxation leads to the lattice constants $a = 9.062$ \AA\ for 1D C$_{60}$ and $a = 9.097$ \AA, $b=9.001$ \AA\ for 2D qTP2 C$_{60}$ respectively.


The band structure calculations using the unscreened hybrid functional give a band gap of 2.46 and 2.13 eV for 1D and 2D C$_{60}$ respectively. The larger band gap of the 1D chain can be attributed to much weakened screening effects as the dimensionality is reduced from a 2D monolayer to a 1D chain. Compared to the band gap difference of about 0.6 eV between monolayer and bulk MoS$_2$\,\cite{Padilha2014}, the relatively small band gap difference between 1D and 2D C$_{60}$ of 0.33 eV is due to their similar dielectric screening. The static dielectric function $\epsilon_{\infty}$ of bulk MoS$_2$ is more than 3.5 times larger than its monolayer counterpart, whereas the $\epsilon_{\infty}$ of monolayer qTP2 C$_{60}$ is approximately 1.5 times the $\epsilon_{\infty}$ of 1D C$_{60}$ chains (for details, see the dielectric function in the Supporting Information). As a result, the band gaps of both 1D and 2D C$_{60}$ are close to the band gap of 2 eV required for efficient photocatalysis\,\cite{LeBahers2014,Wang2019,Brlec2022}.

\begin{figure}
\includegraphics[width = 0.7\linewidth]{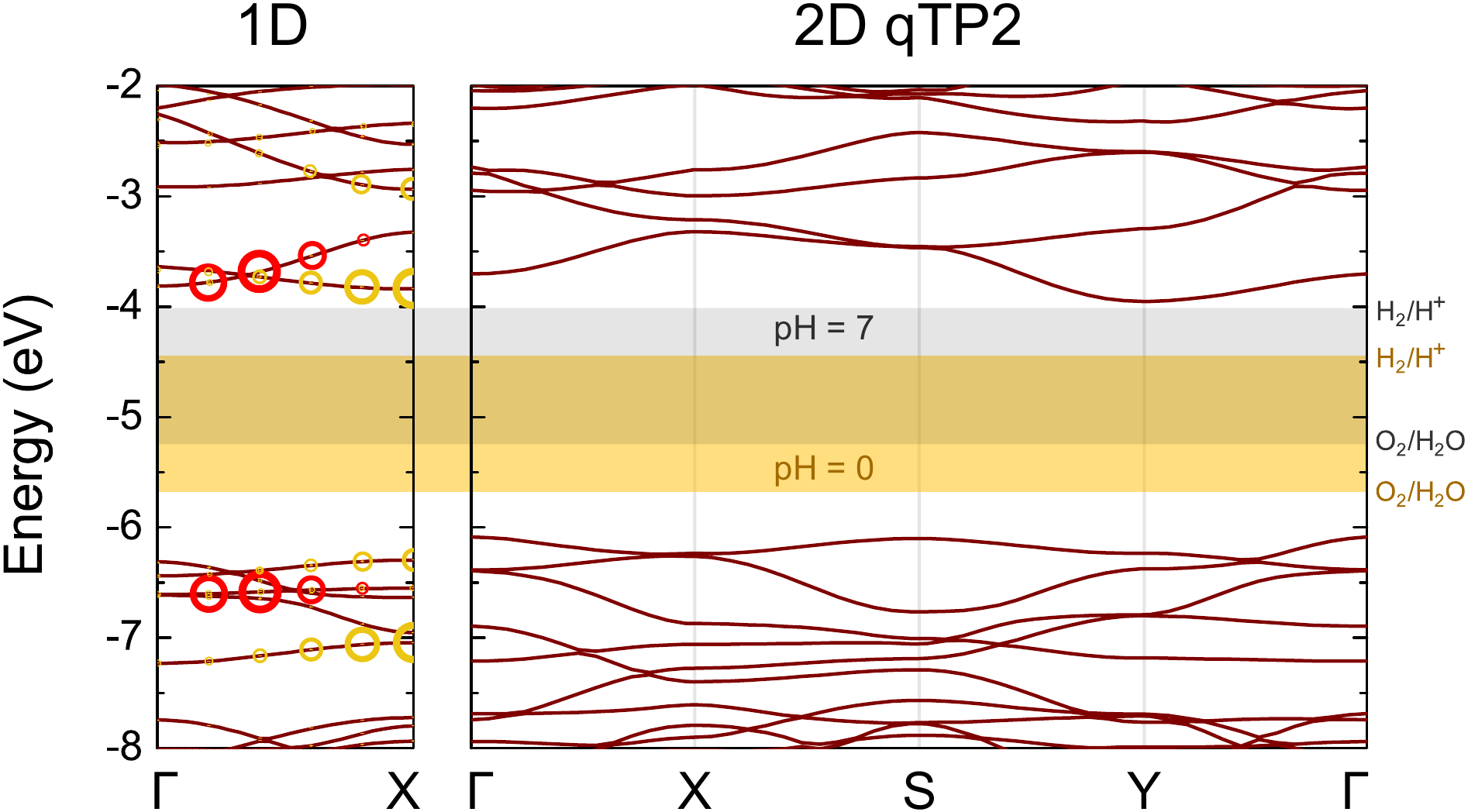}
\caption{Band structures of 1D fullerene chain and 2D fullerene networks computed using the unscreened hybrid functional. The red and yellow circles drawn on the 1D band structures indicate the contributions of the corresponding electron-hole pairs to the two brightest excitons near the band edges. The radii of these circles represent their corresponding oscillator strength.}
\label{band}
\end{figure}

Another requirement for photocatalysts is that the band edges must accommodate the redox potentials involved in the water splitting reaction. The calculated band edges for both systems are shown in Fig.\,\ref{band}, with the vacuum levels calculated by averaging the electrostatic potential as references. 1D C$_{60}$ has a valence band maximum (VBM) at -6.30 eV and a conduction band minimum (CBM) at -3.84 eV with respect to the vacuum level. Both the CBM and VBM lie at the X point in the Brillouin zone, resulting in a direct band gap. On the other hand, 2D qTP2 C$_{60}$ has an indirect band gap with the VBM at $\Gamma$ and the CBM at Y. The VBM of 2D C$_{60}$ at -6.08 eV is 0.22 eV higher than that of 1D C$_{60}$, and its CBM at -3.95 is lower than that of 1D C$_{60}$ by 0.11 eV.


The redox potentials of the relevant half reactions for pH values of 0 and 7 are also plotted in Fig.\,\ref{band}. The CBM of the 1D fullerene polymer is 0.60 eV above the reduction potential of the H$_2$/H$^+$ half reaction at a pH of 0, and it reaches the reduction potential at a pH of 10. In monolayer qTP2 C$_{60}$, the reduction potential of the H$_2$/H$^+$ half reaction is higher than the CBM at a pH of 7. This renders both 1D and 2D C$_{60}$ polymers suitable for the photocatalysis of the hydrogen evolution reaction. However, at a pH of 8, photocatalysis is no longer activated in 2D C$_{60}$, whereas the 1D fullerene polymer can still catalyse the reaction until pH = 10. Therefore the 1D fullerene polymer is able to function as an effective catalyst in a greater range of pH conditions than the 2D qTP2 monolayer.



The first step in photocatalysis is the photoexcitation of the catalysts to generate electron-hole pairs. In this process, it is preferable that the optical absorption is strong enough to generate a large amount of electron-hole pairs, while the excitonic effects are not too strong so the electrons and holes can be separated effectively. The time-dependent Hartree-Fock calculations of excitonic effects reveal the two brightest excitons (with largest oscillator strength) near the band edges in 1D C$_{60}$, and the electron-hole pair contributions are illustrated in Fig.\,\ref{band} as circles, the radii of which reflect the contribution of a particular electron-hole pair to the brightest excitons. The two excitons have eigenenergies of 2.57 and 2.77 eV respectively, corresponding to the red and yellow circles in Fig.\,\ref{band}. The binding energies of the excitons are given by the difference between the independent-particle hybrid-functional eigenenergy and the exciton eigenenergy. The binding energies of the two brightest excitons are 157 and 44 meV respectively. 
Because the holes that contribute the most to the excitons are in much lower valence states than the VBM, carrier thermalisation always leads to the dissociation of the excitons. This process gives rise to effective electron-hole separation, and as a result, both the electrons and holes can proceed to involve themselves in their respective redox reactions.

\begin{figure}
    \centering
    \includegraphics[width=0.5\linewidth]{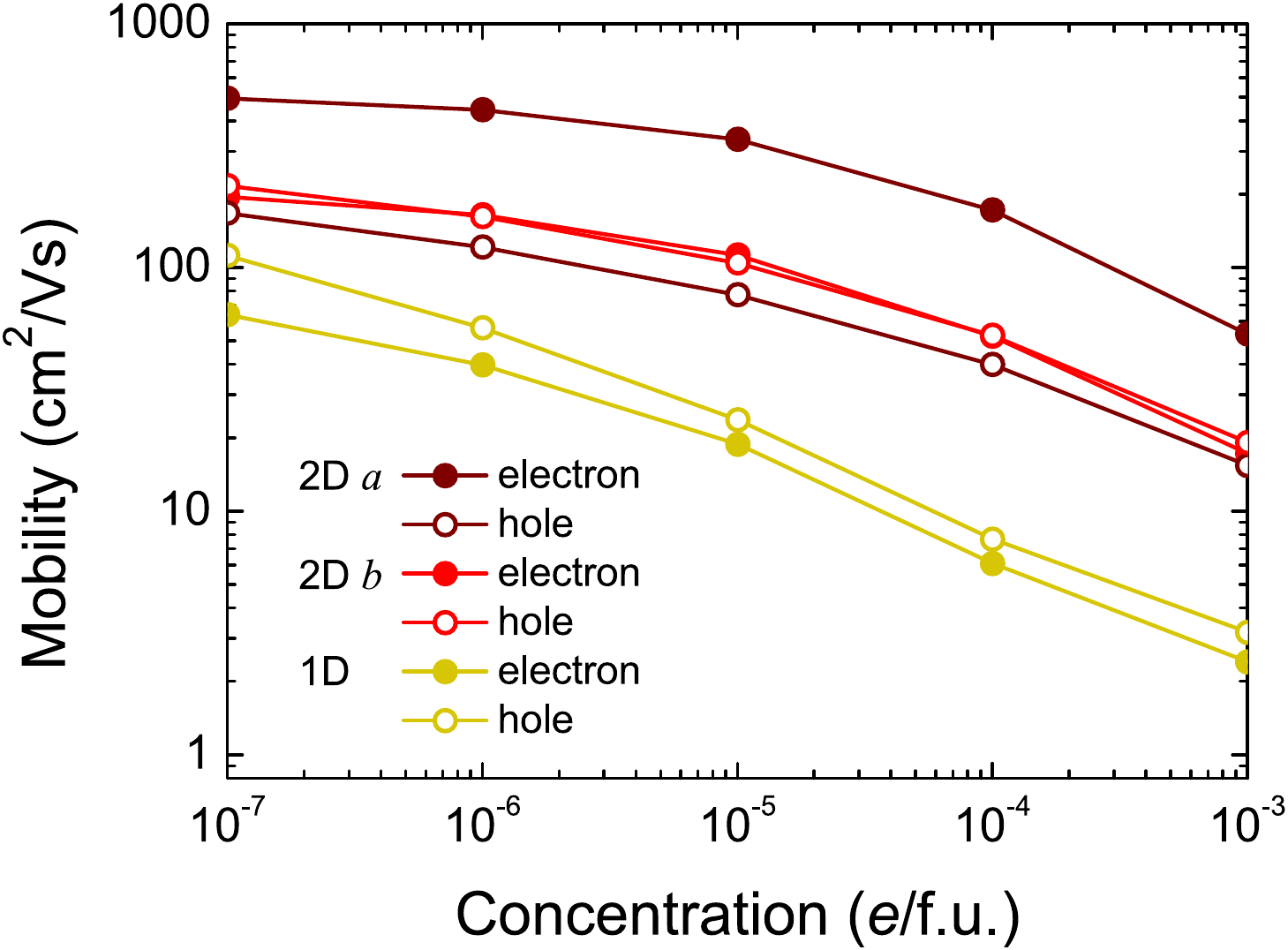}
    \caption{Carrier mobility of 1D and 2D C$_{60}$ at different doping concentrations.}
    \label{mobility}
\end{figure}

To investigate whether the electrons and holes can transfer efficiently after the dissociation of the excitons following the carrier thermalisation, the carrier mobility of 1D and 2D C$_{60}$ is computed at different doping concentrations, as shown in Fig.\,\ref{mobility}. The carrier mobility shows similar trends with increasing concentrations for both systems, as the ionised impurity scattering becomes stronger with more dopants. However, even when the doping concentration is relatively high, i.e. at 10$^{-3}$ $e$/f.u. (charge per unit cell), good carrier mobility can still be obtained, which can be attributed to the delocalised $\pi$ bonds formed near the band edges\,\cite{Peng2022c}.

\begin{figure}
    \centering
    \includegraphics[width=0.7\textwidth]{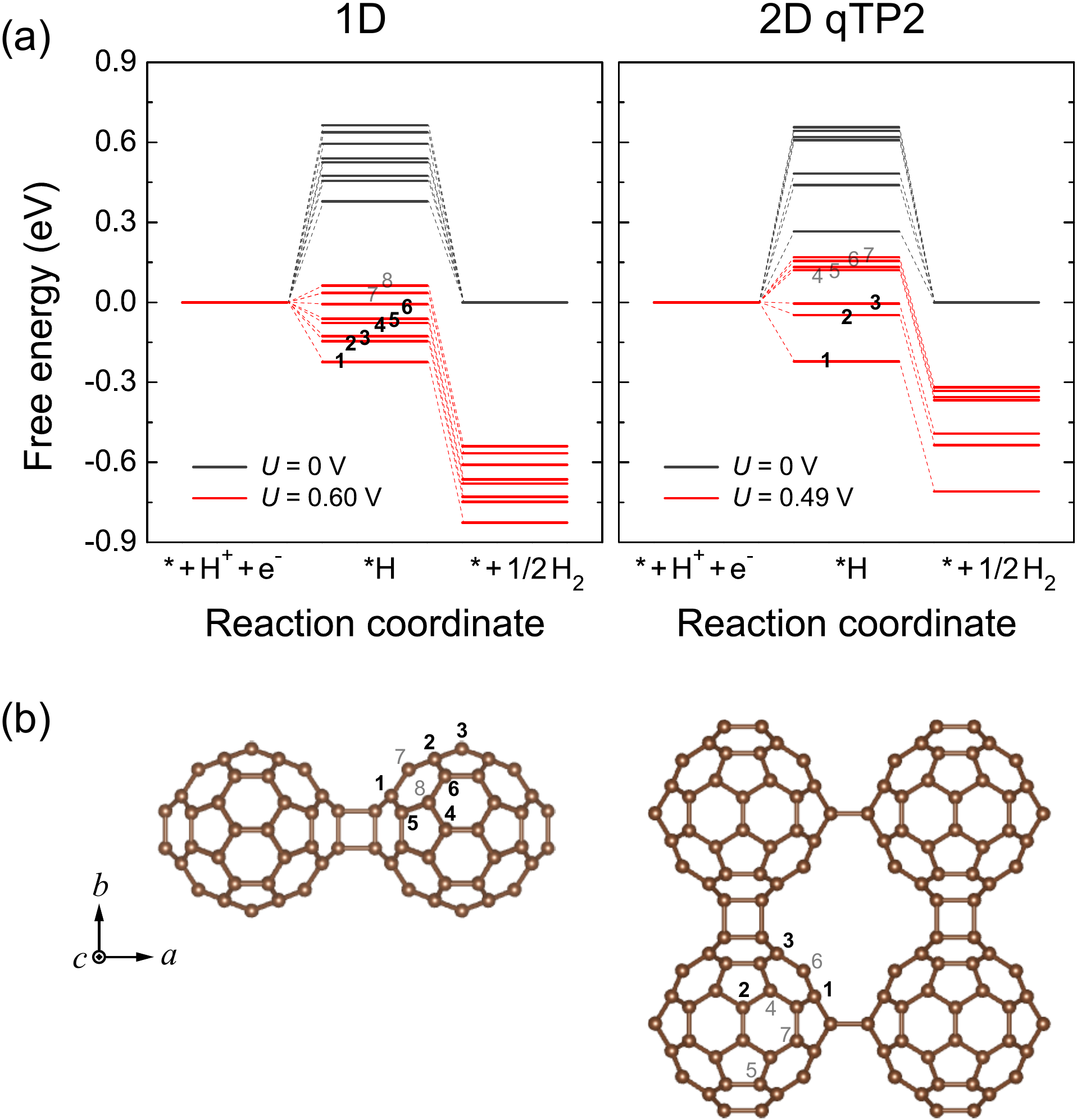}
    \caption{(a) Gibbs free energy changes associated with the hydrogen evolution reaction in the 1D fullerene chain and 2D fullerene networks at a pH of 0 under room temperature, showing the energy barrier posed by the intermediate adsorbate and the effect of photoexcitation in creating a more favourable Gibbs energy in both the intermediates and the products. (b) Adsorption sites for both systems, with the numbers from lower to higher corresponding to free energies of the intermediates from lower to higher.}
    \label{Gibbs}
\end{figure}

As the electrons and holes are separated, they can take part in their respective half reactions. The hydrogen evolution half reaction requires two steps: (1) The photoexcited electron in the conduction band combines with the proton adsorbed on the surface of the catalyst, forming the reaction intermediate $^*$H; (2) The hydrogen atoms adsorbed on the surface of the catalyst form diatomic hydrogen. The energy barrier posed by the intermediates is an important factor in the catalysis process\,\cite{Norskov2004,Rossmeisl2007}. The total change in Gibbs free energy is shown in Fig.\,\ref{Gibbs}(a).


There are two paths shown in Fig.\,\ref{Gibbs}(a) with zero and finite external potential $U$ (in black and red respectively). The $U=0$ reaction path in black corresponds to catalysis without photoexcitation. From the free energy diagram, it can be seen that without photoexcitation to provide an external potential, the production of diatomic hydrogen via the hydrogen evolution reaction described above is not spontaneous, i.e. the energy barrier posed by the intermediate is not negative as is suitable for effective catalysis. 

The other path of a non-zero potential $U$ in red indicates that the photoexcited electron lowers the free energy barrier by $U$. With photoexcitation and the production of electron-hole pairs, electrons in the conduction band can be at high enough energy for the reaction to become spontaneous. During the reaction, the photoexctied electrons combine with protons to form the reaction intermediate $^*$H. In 1D C$_{60}$, an external potential $U = 0.60$ eV is generated by the difference between the CBM and the reduction potential of the H$_2$/H$^+$ half reaction, the free energy change of the intermediate becomes negative for six adsorption sites. In 2D C$_{60}$, $U=0.49$ eV upon photoexcitation, and the free energy change becomes negative for only three adsorption sites at a pH of 0 under room temperature.

For the 1D polymer, the lowest energy intermediate is that for which the hydrogen atom is adsorbed on the carbon atom that is nearest to the [$2+2$] cycloaddition bonds. The energy profiles of the reaction involving each of the symmetry irreducible adsorption sites in the 1D polymer in Fig.\,\ref{Gibbs}(a) correspond to the numbers in Fig.\,\ref{Gibbs}(b), with lower numbers indicating lower Gibbs free energies of the intermediates. Only two adsorption sites among eight symmetry irreducible carbon atoms are not thermodynamically favourable, as marked by 7 and 8 in light grey in Fig.\,\ref{Gibbs}(b). 

For the 2D polymer, only three adsorption sites out of seven are energetically favourable for the hydrogen evolution reaction, with two of them being the nearest neighbouring carbon atoms to the [$2+2$] cycloaddition bonds. Additionally, the remaining four sites, marked by $4-7$ in light grey in Fig.\,\ref{Gibbs}(b), have much larger energy barriers of at least 0.12 eV, which are much larger than the thermal fluctuation energy $k_BT$ at room temperature (0.026 eV) and are therefore not thermally accessible.

The existence of twice the energetically favourable adsorption sites in the 1D chain suggests a greater reaction rate, which renders 1D polymeric fullerene chain a promising candidate as a photocatalyst for the hydrogen evolution reaction. Additionally, 
the 1D chain is thermodynamically more stable than monolayer qTP2 fullerene networks\,\cite{Peng2023}, indicating that the 1D fullerene polymer is a better candidate compared to the 2D qTP2 monolayer as a photocatalyst.



In conclusion, the electronic structures, excitonic effects, transport properties and thermodynamics for the hydrogen evolution reaction of 1D fullerene polymers are investigated in the context of its application as a photocatalyst compared with monolayer fullerene networks. We find that reducing dimensionality from monolayer to chain boosts the photocatalytic performance of polymeric fullerene. The band structures show that the band edges of 1D C$_{60}$ accommodate the reduction potentials of the hydrogen evolution reaction over a wider range of pH conditions than 2D qTP2 C$_{60}$. Time-dependent Hartree-Fock calculations on excitonic effects indicate effective separation of electron-hole pairs in 1D polymeric fullerene chains upon carrier thermalisation, and such separation can further be enhanced by the high carrier mobility. The calculated Gibbs free energy also demonstrates that the 1D fullerene polymer is able to function as a photocatalyst of the hydrogen evolution reaction on two times more adsorption sites than the 2D qTP2 monolayer, as photoexcitation in 1D C$_{60}$ sufficiently reduces the energy barrier posed by the reaction intermediate. Overall, 1D C$_{60}$ can be a much better photocatalyst with higher efficiency and more thermodynamic stability than monolayer C$_{60}$.



\section*{Methods}\label{Methods}

All calculations are performed using \textit{ab initio} methods as implemented by {\sc vasp}\,\cite{Kresse1996,Kresse1996a}. The projector augmented wave (PAW) potential is used\,\cite{Bloechl1994,Kresse1999} with the PBEsol exchange-correlation functional under the generalised gradient approximation (GGA)\,\cite{Perdew2008}. A plane wave cut-off of $800$ eV is used with a \textbf{k}-mesh of $5 \times 1 \times 1$ and $5 \times 5 \times 1$ for 1D and 2D C$_{60}$ respectively. Both the lattice constants and internal coordinates are fully relaxed, until the energy is converged at $10^{-6}$ eV for the self-consistent loop and the force is converged at $10^{-2}$ eV/\AA\ for the geometry optimisation. A 1D fullerene polymer is modelled as a 3D lattice in which the $b$ and $c$ lattice constants are chosen to be 28 \AA\ so that there is negligible interaction between any two 1D fullerene polymers in the 3D lattice. Similarly, for monolayer fullerene networks the $c$ lattice constant is chosen as 24 \AA.

The band structures are calculated using the unscreened hybrid functional, in which the Hartree-Fock and PBEsol exchange energies are mixed in a 1:3 ratio along with the full PBEsol correlation energy\,\cite{Perdew1996a,Adamo1999,Ernzerhof1999}, because it provides better descriptions of the measured electronic and optical band gaps\,\cite{Hou2022,Meirzadeh2023,Peng2022c,Wang2023} (for details, see the band gaps obtained by different levels of theory in the Supporting Information). Following this, the transport properties are calculated by utilising the hybrid-functional eigenenergies and eigenstates in a {\bf k}-mesh of $10\times1 \times 1$ and $8\times8 \times 1$ for 1D and 2D C$_{60}$ respectively, with an interpolation factor of 100. The scattering rates for acoustic deformation potential and ionised impurity scattering are obtained using the {\sc amset} package\,\cite{Ganose2021}, with the elastic tensor coefficients computed using the finite differences method\,\cite{LePage2002,Wu2005} and the static dielectric constant computed using density functional perturbation theory\,\cite{Gajdos2006}. Excitonic effects are considered using the time-dependent Hartree-Fock calculations, which solve the Casida equation using the eigenenergies and wavefunction computed from the unscreened hybrid functional as inputs\,\cite{Casida2012,Sander2017}. A \textbf{k}-mesh of $10 \times 1 \times 1$ and $8 \times 8 \times 1$ is used for 1D and 2D C$_{60}$ respectively with the highest eight valence bands and the lowest eight conduction bands included as the basis to converge the computed exciton eigenenergies within 3 meV.

For the thermodynamics of the hydrogen evolution reaction, geometry optimisation consistently leads to the adsorption of a hydrogen atom on the top site for all symmetry irreducible carbon atoms (except those involved in the [$2+2$] cycloadditional bond that are chemically saturated). To avoid the interactions between hydrogen atoms in neighbouring cells, a supercell of $2\times1 \times 1$ and $2\times2 \times 1$ is used for 1D and 2D C$_{60}$, respectively, with an electronic {\bf k}-point grid of $3 \times 1 \times 1$ and $3 \times 3 \times 1$. All atoms undergo full relaxation, including both the lattice constants and internal atomic coordination. A significant contribution to the free energy of the system is the vibrational energy, and we compute the vibrational energies of both the hydrogen atom and the neighbouring carbon atoms within a radius of 2.5 \AA\ of the adsorbed hydrogen atom. The vibration contribution to the free energy at room temperature is computed by the {\sc vaspkit} package\,\cite{Wang2021b}. The Gibbs free energy of the initial system is calculated using the energy per C$_{60}$ supercell, the energy of half a hydrogen molecule and half the vibrational energy associated with a free H$_2$. Then, the total energy of the intermediate $^*$H is calculated, with the change in vibrational energy between the adsorbate form and pure form included. 




\textbf{Supporting Information:}

Figure\,S1. Dielectric function of 0D, 1D, 2D qTP2 and 3D $Immm$ C$_{60}$, as well as bulk and monolayer MoS$_2$.

Table\,S1. Calculated band gaps (eV) of bilayer and monolayer qHP, monolayer qTP2 and 1D C$_{60}$ using different levels of theory.

\section*{Acknowledgement}

B.P. acknowledges support from the Winton Programme for the Physics of Sustainability, and from Magdalene College Cambridge for a Nevile Research Fellowship. The calculations were performed using resources provided by the Cambridge Service for Data Driven Discovery (CSD3) operated by the University of Cambridge Research Computing Service (www.csd3.cam.ac.uk), provided by Dell EMC and Intel using Tier-2 funding from the Engineering and Physical Sciences Research Council (capital grant EP/T022159/1), and DiRAC funding from the Science and Technology Facilities Council (www.dirac.ac.uk), and by the Cambridge Tier-2 system, operated by the University of Cambridge Research Computing Service (www.hpc.cam.ac.uk) and funded by EPSRC Tier-2 capital grant EP/P020259/1, as well as with computational support from the U.K. Materials and Molecular Modelling Hub, which is partially funded by EPSRC (EP/P020194), for which access is obtained via the UKCP consortium and funded by EPSRC grant ref. EP/P022561/1.


\begin{mcitethebibliography}{57}
\providecommand*\natexlab[1]{#1}
\providecommand*\mciteSetBstSublistMode[1]{}
\providecommand*\mciteSetBstMaxWidthForm[2]{}
\providecommand*\mciteBstWouldAddEndPuncttrue
  {\def\EndOfBibitem{\unskip.}}
\providecommand*\mciteBstWouldAddEndPunctfalse
  {\let\EndOfBibitem\relax}
\providecommand*\mciteSetBstMidEndSepPunct[3]{}
\providecommand*\mciteSetBstSublistLabelBeginEnd[3]{}
\providecommand*\EndOfBibitem{}
\mciteSetBstSublistMode{f}
\mciteSetBstMaxWidthForm{subitem}{(\alph{mcitesubitemcount})}
\mciteSetBstSublistLabelBeginEnd
  {\mcitemaxwidthsubitemform\space}
  {\relax}
  {\relax}

\bibitem[Fujishima and Honda(1972)Fujishima, and Honda]{Fujishima1972}
Fujishima,~A.; Honda,~K. Electrochemical Photolysis of Water at a Semiconductor
  Electrode. \emph{Nature} \textbf{1972}, \emph{238}, 37--38\relax
\mciteBstWouldAddEndPuncttrue
\mciteSetBstMidEndSepPunct{\mcitedefaultmidpunct}
{\mcitedefaultendpunct}{\mcitedefaultseppunct}\relax
\EndOfBibitem
\bibitem[N{\o}rskov et~al.(2004)N{\o}rskov, Rossmeisl, Logadottir, Lindqvist,
  Kitchin, Bligaard, and J{\'o}nsson]{Norskov2004}
N{\o}rskov,~J.~K.; Rossmeisl,~J.; Logadottir,~A.; Lindqvist,~L.;
  Kitchin,~J.~R.; Bligaard,~T.; J{\'o}nsson,~H. Origin of the Overpotential for
  Oxygen Reduction at a Fuel-Cell Cathode. \emph{J. Phys. Chem. B}
  \textbf{2004}, \emph{108}, 17886--17892\relax
\mciteBstWouldAddEndPuncttrue
\mciteSetBstMidEndSepPunct{\mcitedefaultmidpunct}
{\mcitedefaultendpunct}{\mcitedefaultseppunct}\relax
\EndOfBibitem
\bibitem[Rossmeisl et~al.(2007)Rossmeisl, Qu, Zhu, Kroes, and
  N{\o}rskov]{Rossmeisl2007}
Rossmeisl,~J.; Qu,~Z.-W.; Zhu,~H.; Kroes,~G.-J.; N{\o}rskov,~J. Electrolysis of
  water on oxide surfaces. \emph{Journal of Electroanalytical Chemistry}
  \textbf{2007}, \emph{607}, 83--89\relax
\mciteBstWouldAddEndPuncttrue
\mciteSetBstMidEndSepPunct{\mcitedefaultmidpunct}
{\mcitedefaultendpunct}{\mcitedefaultseppunct}\relax
\EndOfBibitem
\bibitem[Zhang et~al.(2007)Zhang, Wang, Zhou, and Xu]{Zhang2007}
Zhang,~L.; Wang,~W.; Zhou,~L.; Xu,~H. Bi$_2$WO$_6$ Nano- and Microstructures:
  Shape Control and Associated Visible-Light-Driven Photocatalytic Activities.
  \emph{Small} \textbf{2007}, \emph{3}, 1618--1625\relax
\mciteBstWouldAddEndPuncttrue
\mciteSetBstMidEndSepPunct{\mcitedefaultmidpunct}
{\mcitedefaultendpunct}{\mcitedefaultseppunct}\relax
\EndOfBibitem
\bibitem[De\'{a}k et~al.(2011)De\'{a}k, Aradi, and Frauenheim]{Deak2011}
De\'{a}k,~P.; Aradi,~B.; Frauenheim,~T. Band Lineup and Charge Carrier
  Separation in Mixed Rutile-Anatase Systems. \emph{J. Phys. Chem. C}
  \textbf{2011}, \emph{115}, 3443--3446\relax
\mciteBstWouldAddEndPuncttrue
\mciteSetBstMidEndSepPunct{\mcitedefaultmidpunct}
{\mcitedefaultendpunct}{\mcitedefaultseppunct}\relax
\EndOfBibitem
\bibitem[Scanlon et~al.(2013)Scanlon, Dunnill, Buckeridge, Shevlin, Logsdail,
  Woodley, Catlow, Powell, Palgrave, Parkin, Watson, Keal, Sherwood, Walsh, and
  Sokol]{Scanlon2013}
Scanlon,~D.~O.; Dunnill,~C.~W.; Buckeridge,~J.; Shevlin,~S.~A.;
  Logsdail,~A.~J.; Woodley,~S.~M.; Catlow,~C. R.~A.; Powell,~M.~J.;
  Palgrave,~R.~G.; Parkin,~I.~P.; Watson,~G.~W.; Keal,~T.~W.; Sherwood,~P.;
  Walsh,~A.; Sokol,~A.~A. {{Band alignment of rutile and anatase TiO$_2$}}.
  \emph{Nature Materials} \textbf{2013}, \emph{12}, 798--801\relax
\mciteBstWouldAddEndPuncttrue
\mciteSetBstMidEndSepPunct{\mcitedefaultmidpunct}
{\mcitedefaultendpunct}{\mcitedefaultseppunct}\relax
\EndOfBibitem
\bibitem[Pfeifer et~al.(2013)Pfeifer, Erhart, Li, Rachut, Morasch, Br\"otz,
  Reckers, Mayer, R\"uhle, Zaban, Mora~Ser\'o, Bisquert, Jaegermann, and
  Klein]{Pfeifer2013}
Pfeifer,~V.; Erhart,~P.; Li,~S.; Rachut,~K.; Morasch,~J.; Br\"otz,~J.;
  Reckers,~P.; Mayer,~T.; R\"uhle,~S.; Zaban,~A.; Mora~Ser\'o,~I.;
  Bisquert,~J.; Jaegermann,~W.; Klein,~A. {{Energy Band Alignment between
  Anatase and Rutile TiO$_2$}}. \emph{J. Phys. Chem. Lett.} \textbf{2013},
  \emph{4}, 4182--4187\relax
\mciteBstWouldAddEndPuncttrue
\mciteSetBstMidEndSepPunct{\mcitedefaultmidpunct}
{\mcitedefaultendpunct}{\mcitedefaultseppunct}\relax
\EndOfBibitem
\bibitem[Ju et~al.(2014)Ju, Sun, Wang, Meng, and Liang]{Ju2014}
Ju,~M.-G.; Sun,~G.; Wang,~J.; Meng,~Q.; Liang,~W. {{Origin of High
  Photocatalytic Properties in the Mixed-Phase TiO$_2$: A First-Principles
  Theoretical Study}}. \emph{ACS Appl. Mater. Interfaces} \textbf{2014},
  \emph{6}, 12885--12892\relax
\mciteBstWouldAddEndPuncttrue
\mciteSetBstMidEndSepPunct{\mcitedefaultmidpunct}
{\mcitedefaultendpunct}{\mcitedefaultseppunct}\relax
\EndOfBibitem
\bibitem[Mi and Weng(2015)Mi, and Weng]{Mi2015}
Mi,~Y.; Weng,~Y. Band Alignment and Controllable Electron Migration between
  Rutile and Anatase TiO$_2$. \emph{Sci. Rep.} \textbf{2015}, \emph{5},
  11482\relax
\mciteBstWouldAddEndPuncttrue
\mciteSetBstMidEndSepPunct{\mcitedefaultmidpunct}
{\mcitedefaultendpunct}{\mcitedefaultseppunct}\relax
\EndOfBibitem
\bibitem[Zhang et~al.(2015)Zhang, Yang, and Dong]{Zhang2015n}
Zhang,~D.; Yang,~M.; Dong,~S. Electric-dipole effect of defects on the energy
  band alignment of rutile and anatase TiO$_2$. \emph{Phys. Chem. Chem. Phys.}
  \textbf{2015}, \emph{17}, 29079--29084\relax
\mciteBstWouldAddEndPuncttrue
\mciteSetBstMidEndSepPunct{\mcitedefaultmidpunct}
{\mcitedefaultendpunct}{\mcitedefaultseppunct}\relax
\EndOfBibitem
\bibitem[De\'{a}k et~al.(2016)De\'{a}k, Kullgren, Aradi, Frauenheim, and
  Kavan]{Deak2016}
De\'{a}k,~P.; Kullgren,~J.; Aradi,~B.; Frauenheim,~T.; Kavan,~L. Water
  splitting and the band edge positions of TiO$_2$. \emph{Electrochimica Acta}
  \textbf{2016}, \emph{199}, 27--34\relax
\mciteBstWouldAddEndPuncttrue
\mciteSetBstMidEndSepPunct{\mcitedefaultmidpunct}
{\mcitedefaultendpunct}{\mcitedefaultseppunct}\relax
\EndOfBibitem
\bibitem[Chiodo et~al.(2010)Chiodo, Garc\'{\i}a-Lastra, Iacomino, Ossicini,
  Zhao, Petek, and Rubio]{Chiodo2010}
Chiodo,~L.; Garc\'{\i}a-Lastra,~J.~M.; Iacomino,~A.; Ossicini,~S.; Zhao,~J.;
  Petek,~H.; Rubio,~A. Self-energy and excitonic effects in the electronic and
  optical properties of ${\text{TiO}}_{2}$ crystalline phases. \emph{Phys. Rev.
  B} \textbf{2010}, \emph{82}, 045207\relax
\mciteBstWouldAddEndPuncttrue
\mciteSetBstMidEndSepPunct{\mcitedefaultmidpunct}
{\mcitedefaultendpunct}{\mcitedefaultseppunct}\relax
\EndOfBibitem
\bibitem[Li et~al.(2020)Li, Wu, and Gao]{Li2020a}
Li,~B.; Wu,~S.; Gao,~X. Theoretical calculation of a TiO$_2$-based
  photocatalyst in the field of water splitting: A review. \emph{Nanotechnology
  Reviews} \textbf{2020}, \emph{9}, 1080--1103\relax
\mciteBstWouldAddEndPuncttrue
\mciteSetBstMidEndSepPunct{\mcitedefaultmidpunct}
{\mcitedefaultendpunct}{\mcitedefaultseppunct}\relax
\EndOfBibitem
\bibitem[Hou et~al.(2022)Hou, Cui, Guan, Wang, Li, Liu, Zhu, and
  Zheng]{Hou2022}
Hou,~L.; Cui,~X.; Guan,~B.; Wang,~S.; Li,~R.; Liu,~Y.; Zhu,~D.; Zheng,~J.
  Synthesis of a monolayer fullerene network. \emph{Nature} \textbf{2022},
  \emph{606}, 507--510\relax
\mciteBstWouldAddEndPuncttrue
\mciteSetBstMidEndSepPunct{\mcitedefaultmidpunct}
{\mcitedefaultendpunct}{\mcitedefaultseppunct}\relax
\EndOfBibitem
\bibitem[Meirzadeh et~al.(2023)Meirzadeh, Evans, Rezaee, Milich, Dionne,
  Darlington, Bao, Bartholomew, Handa, Rizzo, Wiscons, Reza, Zangiabadi,
  Fardian-Melamed, Crowther, Schuck, Basov, Zhu, Giri, Hopkins, Kim,
  Steigerwald, Yang, Nuckolls, and Roy]{Meirzadeh2023}
Meirzadeh,~E. et~al.  A few-layer covalent network of fullerenes. \emph{Nature}
  \textbf{2023}, \emph{613}, 71--76\relax
\mciteBstWouldAddEndPuncttrue
\mciteSetBstMidEndSepPunct{\mcitedefaultmidpunct}
{\mcitedefaultendpunct}{\mcitedefaultseppunct}\relax
\EndOfBibitem
\bibitem[Peng(2022)]{Peng2022c}
Peng,~B. Monolayer Fullerene Networks as Photocatalysts for Overall Water
  Splitting. \emph{J. Am. Chem. Soc.} \textbf{2022}, \emph{144},
  19921--19931\relax
\mciteBstWouldAddEndPuncttrue
\mciteSetBstMidEndSepPunct{\mcitedefaultmidpunct}
{\mcitedefaultendpunct}{\mcitedefaultseppunct}\relax
\EndOfBibitem
\bibitem[Wang et~al.(2023)Wang, Zhang, Wu, Chen, Yang, Lu, and Du]{Wang2023}
Wang,~T.; Zhang,~L.; Wu,~J.; Chen,~M.; Yang,~S.; Lu,~Y.; Du,~P. Few-Layer
  Fullerene Network for Photocatalytic Pure Water Splitting into H$_2$ and
  H$_2$O$_2$. \emph{Angew. Chem. Int. Ed.} \textbf{2023}, \emph{62},
  e202311352\relax
\mciteBstWouldAddEndPuncttrue
\mciteSetBstMidEndSepPunct{\mcitedefaultmidpunct}
{\mcitedefaultendpunct}{\mcitedefaultseppunct}\relax
\EndOfBibitem
\bibitem[Peng(2023)]{Peng2023}
Peng,~B. Stability and Strength of Monolayer Polymeric C$_{60}$. \emph{Nano
  Lett.} \textbf{2023}, \emph{23}, 652--658\relax
\mciteBstWouldAddEndPuncttrue
\mciteSetBstMidEndSepPunct{\mcitedefaultmidpunct}
{\mcitedefaultendpunct}{\mcitedefaultseppunct}\relax
\EndOfBibitem
\bibitem[Xu and Scuseria(1995)Xu, and Scuseria]{Xu1995}
Xu,~C.~H.; Scuseria,~G.~E. Theoretical Predictions for a Two-Dimensional
  Rhombohedral Phase of Solid ${\mathrm{C}}_{60}$. \emph{Phys. Rev. Lett.}
  \textbf{1995}, \emph{74}, 274--277\relax
\mciteBstWouldAddEndPuncttrue
\mciteSetBstMidEndSepPunct{\mcitedefaultmidpunct}
{\mcitedefaultendpunct}{\mcitedefaultseppunct}\relax
\EndOfBibitem
\bibitem[Springborg(1995)]{Springborg1995}
Springborg,~M. Structural and electronic properties of polymeric fullerene
  chains. \emph{Phys. Rev. B} \textbf{1995}, \emph{52}, 2935--2940\relax
\mciteBstWouldAddEndPuncttrue
\mciteSetBstMidEndSepPunct{\mcitedefaultmidpunct}
{\mcitedefaultendpunct}{\mcitedefaultseppunct}\relax
\EndOfBibitem
\bibitem[N{\'u}{\~n}ez-Regueiro et~al.(1995)N{\'u}{\~n}ez-Regueiro, Marques,
  Hodeau, B\'ethoux, and Perroux]{Nunez-Regueiro1995}
N{\'u}{\~n}ez-Regueiro,~M.; Marques,~L.; Hodeau,~J.~L.; B\'ethoux,~O.;
  Perroux,~M. Polymerized Fullerite Structures. \emph{Phys. Rev. Lett.}
  \textbf{1995}, \emph{74}, 278--281\relax
\mciteBstWouldAddEndPuncttrue
\mciteSetBstMidEndSepPunct{\mcitedefaultmidpunct}
{\mcitedefaultendpunct}{\mcitedefaultseppunct}\relax
\EndOfBibitem
\bibitem[Marques et~al.(1996)Marques, Hodeau, N\'u\~nez Regueiro, and
  Perroux]{Marques1996}
Marques,~L.; Hodeau,~J.~L.; N\'u\~nez Regueiro,~M.; Perroux,~M. Pressure and
  temperature diagram of polymerized fullerite. \emph{Phys. Rev. B}
  \textbf{1996}, \emph{54}, R12633--R12636\relax
\mciteBstWouldAddEndPuncttrue
\mciteSetBstMidEndSepPunct{\mcitedefaultmidpunct}
{\mcitedefaultendpunct}{\mcitedefaultseppunct}\relax
\EndOfBibitem
\bibitem[Rao et~al.(1993)Rao, Zhou, Wang, Hager, Holden, Wang, Lee, Bi, Eklund,
  Cornett, Duncan, and Amster]{Rao1993}
Rao,~A.~M.; Zhou,~P.; Wang,~K.-A.; Hager,~G.~T.; Holden,~J.~M.; Wang,~Y.;
  Lee,~W.-T.; Bi,~X.-X.; Eklund,~P.~C.; Cornett,~D.~S.; Duncan,~M.~A.;
  Amster,~I.~J. Photoinduced Polymerization of Solid C$_{60}$ Films.
  \emph{Science} \textbf{1993}, \emph{259}, 955--957\relax
\mciteBstWouldAddEndPuncttrue
\mciteSetBstMidEndSepPunct{\mcitedefaultmidpunct}
{\mcitedefaultendpunct}{\mcitedefaultseppunct}\relax
\EndOfBibitem
\bibitem[Iwasa et~al.(1994)Iwasa, Arima, Fleming, Siegrist, Zhou, Haddon,
  Rothberg, Lyons, Carter, Hebard, Tycko, Dabbagh, Krajewski, Thomas, and
  Yagi]{Iwasa1994}
Iwasa,~Y.; Arima,~T.; Fleming,~R.~M.; Siegrist,~T.; Zhou,~O.; Haddon,~R.~C.;
  Rothberg,~L.~J.; Lyons,~K.~B.; Carter,~H.~L.; Hebard,~A.~F.; Tycko,~R.;
  Dabbagh,~G.; Krajewski,~J.~J.; Thomas,~G.~A.; Yagi,~T. New Phases of C$_{60}$
  Synthesized at High Pressure. \emph{Science} \textbf{1994}, \emph{264},
  1570--1572\relax
\mciteBstWouldAddEndPuncttrue
\mciteSetBstMidEndSepPunct{\mcitedefaultmidpunct}
{\mcitedefaultendpunct}{\mcitedefaultseppunct}\relax
\EndOfBibitem
\bibitem[Eklund et~al.(1995)Eklund, Rao, Zhou, Wang, and Holden]{Eklund1995}
Eklund,~P.; Rao,~A.; Zhou,~P.; Wang,~Y.; Holden,~J. Photochemical
  transformation of C$_{60}$ and C$_{70}$ films. \emph{Thin Solid Films}
  \textbf{1995}, \emph{257}, 185--203\relax
\mciteBstWouldAddEndPuncttrue
\mciteSetBstMidEndSepPunct{\mcitedefaultmidpunct}
{\mcitedefaultendpunct}{\mcitedefaultseppunct}\relax
\EndOfBibitem
\bibitem[Giacalone and Mart\'in(2006)Giacalone, and Mart\'in]{Giacalone2006}
Giacalone,~F.; Mart\'in,~N. Fullerene Polymers: Synthesis and Properties.
  \emph{Chem. Rev.} \textbf{2006}, \emph{106}, 5136--5190\relax
\mciteBstWouldAddEndPuncttrue
\mciteSetBstMidEndSepPunct{\mcitedefaultmidpunct}
{\mcitedefaultendpunct}{\mcitedefaultseppunct}\relax
\EndOfBibitem
\bibitem[{\'A}lvarez~Murga and Hodeau(2015){\'A}lvarez~Murga, and
  Hodeau]{Murga2015}
{\'A}lvarez~Murga,~M.; Hodeau,~J. Structural phase transitions of C$_{60}$
  under high-pressure and high-temperature. \emph{Carbon} \textbf{2015},
  \emph{82}, 381--407\relax
\mciteBstWouldAddEndPuncttrue
\mciteSetBstMidEndSepPunct{\mcitedefaultmidpunct}
{\mcitedefaultendpunct}{\mcitedefaultseppunct}\relax
\EndOfBibitem
\bibitem[Venkateswaran et~al.(1996)Venkateswaran, Sanzi, Krishnappa, Marques,
  Hodeau, N{\'u}{\~n}ez-Regueiro, Rao, and Eklund]{Venkateswaran1996}
Venkateswaran,~U.~D.; Sanzi,~D.; Krishnappa,~J.; Marques,~L.; Hodeau,~J.-L.;
  N{\'u}{\~n}ez-Regueiro,~M.; Rao,~A.~M.; Eklund,~P.~C. Optical Properties of
  Pressure-Polymerized C$_{60}$. \emph{Phys. Stat. Sol. (b)} \textbf{1996},
  \emph{198}, 545--552\relax
\mciteBstWouldAddEndPuncttrue
\mciteSetBstMidEndSepPunct{\mcitedefaultmidpunct}
{\mcitedefaultendpunct}{\mcitedefaultseppunct}\relax
\EndOfBibitem
\bibitem[Rao et~al.(1997)Rao, Eklund, Hodeau, Marques, and
  Nunez-Regueiro]{Rao1997}
Rao,~A.~M.; Eklund,~P.~C.; Hodeau,~J.-L.; Marques,~L.; Nunez-Regueiro,~M.
  Infrared and Raman studies of pressure-polymerized C$_{60}$s. \emph{Phys.
  Rev. B} \textbf{1997}, \emph{55}, 4766--4773\relax
\mciteBstWouldAddEndPuncttrue
\mciteSetBstMidEndSepPunct{\mcitedefaultmidpunct}
{\mcitedefaultendpunct}{\mcitedefaultseppunct}\relax
\EndOfBibitem
\bibitem[Forr\'o and Mih\'aly(2001)Forr\'o, and Mih\'aly]{Forro2001}
Forr\'o,~L.; Mih\'aly,~L. Electronic properties of doped fullerenes. \emph{Rep.
  Prog. Phys.} \textbf{2001}, \emph{64}, 649\relax
\mciteBstWouldAddEndPuncttrue
\mciteSetBstMidEndSepPunct{\mcitedefaultmidpunct}
{\mcitedefaultendpunct}{\mcitedefaultseppunct}\relax
\EndOfBibitem
\bibitem[Makarova(2001)]{Makarova2001}
Makarova,~T.~L. Electrical and optical properties of pristine and polymerized
  fullerenes. \emph{Semiconductors} \textbf{2001}, \emph{35}, 243--278\relax
\mciteBstWouldAddEndPuncttrue
\mciteSetBstMidEndSepPunct{\mcitedefaultmidpunct}
{\mcitedefaultendpunct}{\mcitedefaultseppunct}\relax
\EndOfBibitem
\bibitem[Sun et~al.(2005)Sun, Liang, Yang, and Gao]{Sun2005}
Sun,~J.; Liang,~W.; Yang,~J.; Gao,~J. A comparative investigation on electronic
  structures and optical properties of C$_{60}$ polymers, periodic fulleriods
  and carbon nanotubes. \emph{Journal of Molecular Structure: THEOCHEM}
  \textbf{2005}, \emph{755}, 105--111\relax
\mciteBstWouldAddEndPuncttrue
\mciteSetBstMidEndSepPunct{\mcitedefaultmidpunct}
{\mcitedefaultendpunct}{\mcitedefaultseppunct}\relax
\EndOfBibitem
\bibitem[Belosludov et~al.(2003)Belosludov, Inerbaev, Belosludov, and
  Kawazoe]{Belosludov2003}
Belosludov,~V.~R.; Inerbaev,~T.~M.; Belosludov,~R.~V.; Kawazoe,~Y. Polaron in a
  one-dimensional C$_{60}$ crystal. \emph{Phys. Rev. B} \textbf{2003},
  \emph{67}, 155410\relax
\mciteBstWouldAddEndPuncttrue
\mciteSetBstMidEndSepPunct{\mcitedefaultmidpunct}
{\mcitedefaultendpunct}{\mcitedefaultseppunct}\relax
\EndOfBibitem
\bibitem[Belosludov et~al.(2006)Belosludov, Inerbaev, Belosludov, Kawazoe, and
  Kudoh]{Belosludov2006}
Belosludov,~V.; Inerbaev,~T.; Belosludov,~R.; Kawazoe,~Y.; Kudoh,~J.
  Theoretical study of polarons and self-trapped excited states in
  one-dimensional C$_{60}$ crystal. \emph{Computational Materials Science}
  \textbf{2006}, \emph{36}, 17--25\relax
\mciteBstWouldAddEndPuncttrue
\mciteSetBstMidEndSepPunct{\mcitedefaultmidpunct}
{\mcitedefaultendpunct}{\mcitedefaultseppunct}\relax
\EndOfBibitem
\bibitem[Chauvet et~al.(1994)Chauvet, Oszl\`anyi, Forro, Stephens, Tegze,
  Faigel, and J\`anossy]{Chauvet1994}
Chauvet,~O.; Oszl\`anyi,~G.; Forro,~L.; Stephens,~P.~W.; Tegze,~M.; Faigel,~G.;
  J\`anossy,~A. Quasi-one-dimensional electronic structure in orthorhombic
  ${\mathrm{RbC}}_{60}$. \emph{Phys. Rev. Lett.} \textbf{1994}, \emph{72},
  2721--2724\relax
\mciteBstWouldAddEndPuncttrue
\mciteSetBstMidEndSepPunct{\mcitedefaultmidpunct}
{\mcitedefaultendpunct}{\mcitedefaultseppunct}\relax
\EndOfBibitem
\bibitem[Stephens et~al.(1994)Stephens, Bortel, Faigel, Tegze, J\'anossy,
  Pekker, Oszlanyi, and Forr\'o]{Stephens1994}
Stephens,~P.~W.; Bortel,~G.; Faigel,~G.; Tegze,~M.; J\'anossy,~A.; Pekker,~S.;
  Oszlanyi,~G.; Forr\'o,~L. Polymeric fullerene chains in RbC$_{60}$ and
  KC$_{60}$. \emph{Nature} \textbf{1994}, \emph{370}, 636--639\relax
\mciteBstWouldAddEndPuncttrue
\mciteSetBstMidEndSepPunct{\mcitedefaultmidpunct}
{\mcitedefaultendpunct}{\mcitedefaultseppunct}\relax
\EndOfBibitem
\bibitem[Gunnarsson(1997)]{Gunnarsson1997}
Gunnarsson,~O. Superconductivity in fullerides. \emph{Rev. Mod. Phys.}
  \textbf{1997}, \emph{69}, 575--606\relax
\mciteBstWouldAddEndPuncttrue
\mciteSetBstMidEndSepPunct{\mcitedefaultmidpunct}
{\mcitedefaultendpunct}{\mcitedefaultseppunct}\relax
\EndOfBibitem
\bibitem[Huq et~al.(2001)Huq, Stephens, Bendele, and Ibberson]{Huq2001}
Huq,~A.; Stephens,~P.; Bendele,~G.~M.; Ibberson,~R. Polymeric fullerene chains
  in RbC$_{60}$ and KC$_{60}$. \emph{Chemical Physics Letters} \textbf{2001},
  \emph{347}, 13--22\relax
\mciteBstWouldAddEndPuncttrue
\mciteSetBstMidEndSepPunct{\mcitedefaultmidpunct}
{\mcitedefaultendpunct}{\mcitedefaultseppunct}\relax
\EndOfBibitem
\bibitem[Le~Bahers et~al.(2014)Le~Bahers, R\'{e}rat, and Sautet]{LeBahers2014}
Le~Bahers,~T.; R\'{e}rat,~M.; Sautet,~P. Semiconductors Used in Photovoltaic
  and Photocatalytic Devices: Assessing Fundamental Properties from DFT.
  \emph{J. Phys. Chem. C} \textbf{2014}, \emph{118}, 5997--6008\relax
\mciteBstWouldAddEndPuncttrue
\mciteSetBstMidEndSepPunct{\mcitedefaultmidpunct}
{\mcitedefaultendpunct}{\mcitedefaultseppunct}\relax
\EndOfBibitem
\bibitem[Wang et~al.(2019)Wang, Li, and Domen]{Wang2019}
Wang,~Z.; Li,~C.; Domen,~K. Recent developments in heterogeneous photocatalysts
  for solar-driven overall water splitting. \emph{Chem. Soc. Rev.}
  \textbf{2019}, \emph{48}, 2109--2125\relax
\mciteBstWouldAddEndPuncttrue
\mciteSetBstMidEndSepPunct{\mcitedefaultmidpunct}
{\mcitedefaultendpunct}{\mcitedefaultseppunct}\relax
\EndOfBibitem
\bibitem[Brlec et~al.(2022)Brlec, Kavanagh, Savory, and Scanlon]{Brlec2022}
Brlec,~K.; Kavanagh,~S.~R.; Savory,~C.~N.; Scanlon,~D.~O. Understanding the
  Photocatalytic Activity of La$_5$Ti$_2$AgS$_5$O$_7$ and
  La$_5$Ti$_2$CuS$_5$O$_7$ for Green Hydrogen Production: Computational
  Insights. \emph{ACS Appl. Energy Mater.} \textbf{2022}, \emph{5},
  1992--2001\relax
\mciteBstWouldAddEndPuncttrue
\mciteSetBstMidEndSepPunct{\mcitedefaultmidpunct}
{\mcitedefaultendpunct}{\mcitedefaultseppunct}\relax
\EndOfBibitem
\bibitem[Kresse and Furthm\"uller(1996)Kresse, and Furthm\"uller]{Kresse1996}
Kresse,~G.; Furthm\"uller,~J. {Efficient iterative schemes for \textit{ab
  initio} total-energy calculations using a plane-wave basis set}. \emph{Phys.
  Rev. B} \textbf{1996}, \emph{54}, 11169--11186\relax
\mciteBstWouldAddEndPuncttrue
\mciteSetBstMidEndSepPunct{\mcitedefaultmidpunct}
{\mcitedefaultendpunct}{\mcitedefaultseppunct}\relax
\EndOfBibitem
\bibitem[Kresse and Furthm\"uller(1996)Kresse, and Furthm\"uller]{Kresse1996a}
Kresse,~G.; Furthm\"uller,~J. Efficiency of ab-initio total energy calculations
  for metals and semiconductors using a plane-wave basis set.
  \emph{Computational Materials Science} \textbf{1996}, \emph{6}, 15 --
  50\relax
\mciteBstWouldAddEndPuncttrue
\mciteSetBstMidEndSepPunct{\mcitedefaultmidpunct}
{\mcitedefaultendpunct}{\mcitedefaultseppunct}\relax
\EndOfBibitem
\bibitem[Bl\"ochl(1994)]{Bloechl1994}
Bl\"ochl,~P.~E. Projector augmented-wave method. \emph{Phys. Rev. B}
  \textbf{1994}, \emph{50}, 17953--17979\relax
\mciteBstWouldAddEndPuncttrue
\mciteSetBstMidEndSepPunct{\mcitedefaultmidpunct}
{\mcitedefaultendpunct}{\mcitedefaultseppunct}\relax
\EndOfBibitem
\bibitem[Kresse and Joubert(1999)Kresse, and Joubert]{Kresse1999}
Kresse,~G.; Joubert,~D. From ultrasoft pseudopotentials to the projector
  augmented-wave method. \emph{Phys. Rev. B} \textbf{1999}, \emph{59},
  1758--1775\relax
\mciteBstWouldAddEndPuncttrue
\mciteSetBstMidEndSepPunct{\mcitedefaultmidpunct}
{\mcitedefaultendpunct}{\mcitedefaultseppunct}\relax
\EndOfBibitem
\bibitem[Perdew et~al.(2008)Perdew, Ruzsinszky, Csonka, Vydrov, Scuseria,
  Constantin, Zhou, and Burke]{Perdew2008}
Perdew,~J.~P.; Ruzsinszky,~A.; Csonka,~G.~I.; Vydrov,~O.~A.; Scuseria,~G.~E.;
  Constantin,~L.~A.; Zhou,~X.; Burke,~K. {{Restoring the Density-Gradient
  Expansion for Exchange in Solids and Surfaces}}. \emph{Phys. Rev. Lett.}
  \textbf{2008}, \emph{100}, 136406\relax
\mciteBstWouldAddEndPuncttrue
\mciteSetBstMidEndSepPunct{\mcitedefaultmidpunct}
{\mcitedefaultendpunct}{\mcitedefaultseppunct}\relax
\EndOfBibitem
\bibitem[Perdew et~al.(1996)Perdew, Ernzerhof, and Burke]{Perdew1996a}
Perdew,~J.~P.; Ernzerhof,~M.; Burke,~K. Rationale for mixing exact exchange
  with density functional approximations. \emph{J. Chem. Phys.} \textbf{1996},
  \emph{105}, 9982--9985\relax
\mciteBstWouldAddEndPuncttrue
\mciteSetBstMidEndSepPunct{\mcitedefaultmidpunct}
{\mcitedefaultendpunct}{\mcitedefaultseppunct}\relax
\EndOfBibitem
\bibitem[Adamo and Barone(1999)Adamo, and Barone]{Adamo1999}
Adamo,~C.; Barone,~V. Toward reliable density functional methods without
  adjustable parameters: The PBE0 model. \emph{J. Chem. Phys.} \textbf{1999},
  \emph{110}, 6158--6170\relax
\mciteBstWouldAddEndPuncttrue
\mciteSetBstMidEndSepPunct{\mcitedefaultmidpunct}
{\mcitedefaultendpunct}{\mcitedefaultseppunct}\relax
\EndOfBibitem
\bibitem[Ernzerhof and Scuseria(1999)Ernzerhof, and Scuseria]{Ernzerhof1999}
Ernzerhof,~M.; Scuseria,~G.~E. Assessment of the Perdew-Burke-Ernzerhof
  exchange-correlation functional. \emph{J. Chem. Phys.} \textbf{1999},
  \emph{110}, 5029--5036\relax
\mciteBstWouldAddEndPuncttrue
\mciteSetBstMidEndSepPunct{\mcitedefaultmidpunct}
{\mcitedefaultendpunct}{\mcitedefaultseppunct}\relax
\EndOfBibitem
\bibitem[Ganose et~al.(2021)Ganose, Park, Faghaninia, Woods-Robinson, Persson,
  and Jain]{Ganose2021}
Ganose,~A.~M.; Park,~J.; Faghaninia,~A.; Woods-Robinson,~R.; Persson,~K.~A.;
  Jain,~A. Efficient calculation of carrier scattering rates from first
  principles. \emph{Nature Communications} \textbf{2021}, \emph{12},
  2222--\relax
\mciteBstWouldAddEndPuncttrue
\mciteSetBstMidEndSepPunct{\mcitedefaultmidpunct}
{\mcitedefaultendpunct}{\mcitedefaultseppunct}\relax
\EndOfBibitem
\bibitem[Le~Page and Saxe(2002)Le~Page, and Saxe]{LePage2002}
Le~Page,~Y.; Saxe,~P. Symmetry-general least-squares extraction of elastic data
  for strained materials from \textit{ab initio} calculations of stress.
  \emph{Phys. Rev. B} \textbf{2002}, \emph{65}, 104104\relax
\mciteBstWouldAddEndPuncttrue
\mciteSetBstMidEndSepPunct{\mcitedefaultmidpunct}
{\mcitedefaultendpunct}{\mcitedefaultseppunct}\relax
\EndOfBibitem
\bibitem[Wu et~al.(2005)Wu, Vanderbilt, and Hamann]{Wu2005}
Wu,~X.; Vanderbilt,~D.; Hamann,~D.~R. Systematic treatment of displacements,
  strains, and electric fields in density-functional perturbation theory.
  \emph{Phys. Rev. B} \textbf{2005}, \emph{72}, 035105\relax
\mciteBstWouldAddEndPuncttrue
\mciteSetBstMidEndSepPunct{\mcitedefaultmidpunct}
{\mcitedefaultendpunct}{\mcitedefaultseppunct}\relax
\EndOfBibitem
\bibitem[Gajdo\v{s} et~al.(2006)Gajdo\v{s}, Hummer, Kresse, Furthm\"uller, and
  Bechstedt]{Gajdos2006}
Gajdo\v{s},~M.; Hummer,~K.; Kresse,~G.; Furthm\"uller,~J.; Bechstedt,~F. Linear
  optical properties in the projector-augmented wave methodology. \emph{Phys.
  Rev. B} \textbf{2006}, \emph{73}, 045112\relax
\mciteBstWouldAddEndPuncttrue
\mciteSetBstMidEndSepPunct{\mcitedefaultmidpunct}
{\mcitedefaultendpunct}{\mcitedefaultseppunct}\relax
\EndOfBibitem
\bibitem[Casida and Huix-Rotllant(2012)Casida, and Huix-Rotllant]{Casida2012}
Casida,~M.; Huix-Rotllant,~M. Progress in Time-Dependent Density-Functional
  Theory. \emph{Annu. Rev. Phys. Chem.} \textbf{2012}, \emph{63},
  287--323\relax
\mciteBstWouldAddEndPuncttrue
\mciteSetBstMidEndSepPunct{\mcitedefaultmidpunct}
{\mcitedefaultendpunct}{\mcitedefaultseppunct}\relax
\EndOfBibitem
\bibitem[Sander and Kresse(2017)Sander, and Kresse]{Sander2017}
Sander,~T.; Kresse,~G. Macroscopic dielectric function within time-dependent
  density functional theory--Real time evolution versus the Casida approach.
  \emph{J. Chem. Phys.} \textbf{2017}, \emph{146}, 064110\relax
\mciteBstWouldAddEndPuncttrue
\mciteSetBstMidEndSepPunct{\mcitedefaultmidpunct}
{\mcitedefaultendpunct}{\mcitedefaultseppunct}\relax
\EndOfBibitem
\bibitem[Wang et~al.(2021)Wang, Xu, Liu, Tang, and Geng]{Wang2021b}
Wang,~V.; Xu,~N.; Liu,~J.-C.; Tang,~G.; Geng,~W.-T. VASPKIT: A user-friendly
  interface facilitating high-throughput computing and analysis using VASP
  code. \emph{Computer Physics Communications} \textbf{2021}, \emph{267},
  108033\relax
\mciteBstWouldAddEndPuncttrue
\mciteSetBstMidEndSepPunct{\mcitedefaultmidpunct}
{\mcitedefaultendpunct}{\mcitedefaultseppunct}\relax
\EndOfBibitem
\end{mcitethebibliography}

\providecommand*\mcitethebibliography{\thebibliography}
\csname @ifundefined\endcsname{endmcitethebibliography}
  {\let\endmcitethebibliography\endthebibliography}{}

\end{document}